\documentclass{llncs}

\usepackage{array}
\usepackage{graphicx}
\usepackage{amsmath}
\usepackage{amsfonts}
\usepackage{hyperref}
\usepackage{xcolor}
\usepackage[utf8]{inputenc}

\usepackage{hyperref}
\usepackage{soul}
\usepackage{float}

\title{Automatic Tissue Segmentation with Deep Learning in Patients
  with Congenital or Acquired Distortion of Brain Anatomy}

\author{Gabriele Amorosino\inst{1,2}, Denis Peruzzo\inst{3}, Pietro
  Astolfi\inst{1,4}, Daniela Redaelli\inst{3}, Paolo
  Avesani\inst{1,2}, Filippo Arrigoni\inst{3} and Emanuele
  Olivetti\inst{1,2}}

\institute{NeuroInformatics Laboratory (NILab), Bruno Kessler
  Foundation, Trento, Italy 
  \and 
  Center for Mind and Brain Sciences
  (CIMeC), University of Trento, Italy
  \and
  Neuroimaging Lab, Scientific Institute IRCCS Eugenio Medea, \\
  Bosisio Parini (Lecco), Italy
  \and
  PAVIS, Italian Institute of Technology (IIT), Genova, Italy\\
  \email{olivetti@fbk.eu \quad gamorosino@fbk.eu}
}

\begin{document}

\maketitle

\begin{abstract}
  Brains with complex distortion of cerebral anatomy present several
  challenges to automatic tissue segmentation methods of T1-weighted
  MR images. First, the very high variability in the morphology of the
  tissues can be incompatible with the prior knowledge embedded within
  the algorithms. Second, the availability of MR images of distorted
  brains is very scarce, so the methods in the literature have not
  addressed such cases so far. In this work, we present the first
  evaluation of state-of-the-art automatic tissue segmentation
  pipelines on T1-weighted images of brains with different severity of
  congenital or acquired brain distortion. We compare traditional
  pipelines and a deep learning model, i.e. a 3D U-Net trained on
  normal-appearing brains. Unsurprisingly, traditional pipelines
  completely fail to segment the tissues with strong anatomical
  distortion. Surprisingly, the 3D U-Net provides useful segmentations
  that can be a valuable starting point for manual refinement by
  experts/neuroradiologists.
\end{abstract}

\section{Introduction}
\label{sec:introduction}


Accurate segmentation of brain structural MR images into different
tissues, like white matter (WM), gray matter (GM) and cerebrospinal
fluid (CSF), is of primary interest for clinical and neuroscientific
applications, such as volume quantification, cortical thickness
analysis and bundle analysis.

Since manual segmentation of the brain tissues is extremely time
consuming it is usually performed by means of well-established
automated tools, such as FSL \cite{jenkinson_fsl_2012}, SPM
\cite{ashburner_voxel-based_2000}, FreeSurfer
\cite{dale_cortical_1999} and ANTs
\cite{tustison_large-scale_2014}. Typically, these tools obtain
excellent quality of segmentation in normal-appearing brains.

More recently, brain tissue segmentation has been addressed by deep
learning algorithms like convolutional neural networks (CNNs)
\cite{moeskops_automatic_2016,rajchl_neuronet_2018,spalletta_convolutional_2018,yogananda_deep_2019},
applied directly to T1 or T2 weighted MR images. The quality of
segmentation obtained by such methods is again excellent and the
computational cost, once trained, is usually greatly reduced with
respect to traditional pipelines.

Especially in children, many congenital (e.g. malformations, huge
arachnoid cysts) or acquired (e.g. severe hydrocephalus,
encephalomalacia due to perinatal injuries) conditions can cause
complex modifications of cerebral anatomy that alter the structural
and spatial relationship among different brain
structures. Automatically segmenting such brains presents multiple
challenges mainly due to the high variability of the morphology
together with the scarce availability of data. Moreover, the prior
knowledge encoded in automated pipelines, or the set of images used to
train segmentation algorithms, do not cover such cases.

In this work, for the first time, we present results of different
well-established brain tissue segmentation pipelines on T1 images of
malformed and highly distorted brains in the pediatric
age. Unsurprisingly, we observe that the quality of segmentation is
highly variable with traditional pipelines, and it fails when the
complexity of the brain distortion and the severity of the
malformation are high.

Moreover, as a major contribution, we show the results of a CNN for
segmentation of medical images, namely the 3D U-Net
\cite{ourselin_3d_2016,ronneberger_u-net:_2015}, trained on over 800
pediatric subjects with \emph{normal brain}. Surprisingly, the 3D
U-Net segments brains with moderate and severe distortion of brain
anatomy either accurately or at least to a sufficient level to
consider manual refinement by expert radiologists.

The evaluation study of the U-Net presented here is first conducted on
a large sample of normal brain images, as a sanity check, to
quantitatively assess that the specific implementation and training
procedure reaches state-of-the-art quality of segmentation. The
evaluation on distorted brains is instead qualitative, because of the
small sample available due to the rarity of the condition, as well as
the current lack of gold standard segmentation.

If confirmed by future and more extensive studies, this result opens
the way to CNNs-based tissue segmentation methods in applications,
even in the case of malformed and highly distorted brains. From the
methodological point of view, CNNs show a much higher degree of
flexibility in tissue segmentation than well-established pipelines,
despite being trained on segmentations obtained from those same
pipelines.

\section{Materials}
\label{sec:materials}
We assembled a dataset of over 900 MR images from subjects and
patients in the pediatric age, divided into two parts: normal brains
and distorted brains. The first part consists of 570 T1-w images from
public databases
(C-MIND\footnote{\url{https://research.cchmc.org/c-mind}, NIH contract
  \#s HHSN275200900018C.},
NIMH\footnote{\url{http://pediatricmri.nih.gov}}) and 334 T1-w images
acquired in-house by the authors during clinical activity at 
IRCCS Eugenio Medea (Italy).
The second part comprises 21 patients, again acquired at 
IRCCS E.Medea.

\subsection{Normal Brains}
\label{sec:healthy}
\begin{itemize}
\item \textbf{Public Databases.}
  \begin{itemize}
  \item T1-w images from 207 healthy subjects of the C-MIND database
    (165 from CCHMC site, with average age 8.9 (SD=5.0) and 42 from
    UCLA site, with average age 7.6 (SD=3.8)) both with 3D MPRAGE MRI
    3T scan sequence (TR=8000ms, TE=3.7ms, Flip angle=8$^o$,
    FOV=256x224x160, voxel spacing=1x1x1 mm$^3$).
  \item T1-w images from 363 healthy subjects (average age 10.7
    (SD=6.0)) from the \emph{Pediatric MRI} database of NIMH Data
    Archive \cite{evans_nih_2006}, MRI 1.5T scanner with two different
    sequences:
    \begin{itemize}
    \item 284 subjects acquired with a 3D T1-w Sequence 3D RF-spoiled
      gradient echo sequence (TR 22$-$25 ms, TE=10$-$11 ms, Excitation
      pulse =30$^o$, Refocusing pulse 180$^o$, FOV =
      160$-$180x256x256, voxel spacing=1x1x1 mm$^3$)
    \item 79 subjects acquired with a T1-w Sequence Spin echo (TR =
      500 ms, TE = 12 ms, Flip angle= 90$^o$, Refocusing pulse=180$^o$
      Field of view = 192x256x66, voxel spacing=1x1x3 mm$^3$)
    \end{itemize}
  \end{itemize}
\item \textbf{In-House Database: IRCCS E.MEDEA.} 
  T1-w images from 334 subjects with normal brain, acquired in-house
  with average age 10.6 (SD=5.2) and 3D T1-w MPRAGE 3T MRI scan
  sequence (TR=8000ms, TE=4ms, Flip angle=8$^o$, FOV=256x256x170,
  voxel spacing=1x1x1 mm$^3$).
\end{itemize}

\subsection{Distorted Brains}
\label{sec:distorted}
Images from two groups of patients acquired in-house at 
IRCSS E.Medea
with the same MR and scan sequence described above. In detail:
\begin{itemize}
\item \textbf{Agenesis of Corpus Callosum.} 12 patients with agenesis
  of corpus callosum (ACC) (average age 5.8 (SD=5.2)). Callosal
  agenesis is characterized by colpocephaly, parallel ventricles,
  presence of Probst bundles and upward extension of third
  ventricle. See some cases in Figure \ref{fig:acc}.
\item \textbf{Complex distortions.} 9 patients (average age 7.9
  (SD=3.2)) with severe parenchymal distortion related to complex
  malformations (4 cases), parenchymal poroencephalic lesions and
  severe hydrocephalus (4 cases) and massive arachnoid cyst (1
  case). See some cases in Figure \ref{fig:complex}.
\end{itemize}

\subsection{Pre-processing and Reference Tissue Segmentation}
\label{sec:preprocessing}
All T1-w images received bias field correction
(\texttt{N4BiasFieldCorrection}) and AC-PC alignment. The reference
segmentation of \emph{normal brains} was performed with the
\texttt{AntsCorticalThickness.sh} script of ANTs
\cite{tustison_large-scale_2014} with the PTBP (pediatric) prior
\cite{avants_pediatric_2015}, resulting in a 3D mask with 7 labels (6
tissues): background, Cerebrospinal fluid (CSF), Gray matter (GM),
White matter (WM), Deep gray matter (DGM), Trunk and Cerebellum. The
results of the segmentations were visually assessed by two experts in
pediatric neuroimaging.

\section{Methods}
\label{sec:methods}


A standard whole-brain 3D U-Net model
\cite{ourselin_3d_2016,ronneberger_u-net:_2015} was implemented to
predict the brain masks of 6 tissues plus background from T1-w
images. The architecture of the network consists of 3 main parts: the
contraction path, the bottleneck and the expansion path. The input and
the output layer are for 256x256x256 isotropic 3D images, resolution
to which all input images are initially resampled. The adopted
architecture, described below, has minor changes with respect to the
literature to reduce the memory footprint and to fit the GPU used
during the experiments reported in Section \ref{sec:experiments}.

The contraction path consists of 4 blocks, each with two convolutional
layers (3x3x3 kernel followed by ReLU) followed by MaxPooling (2x2x2
filter, stride 2x2x2) for downsampling. The size of the downsampling
over the blocks is
$256 \rightarrow 128 \rightarrow 64 \rightarrow 32 \rightarrow 16$,
while the application of convolutional layers produces an increasing
number of feature maps over the blocks,
$12 \rightarrow 24 \rightarrow 48 \rightarrow 96$.  The bottleneck
consists of 2 convolutional layers both of 192 filters (3x3x3 kernel
then ReLU), followed by a dropout layer (dropout rate: 0.15) used to
prevent overfitting.  Similarly to the contraction path, the expansion
path consists of 4 blocks, each with transposed convolution for
upsampling, followed by concatenation with contraction features (skip
connected) and two convolutions.
The upsampling size over the blocks is the opposite of the
downsampling, i.e. from $16$ to $256$.  All convolutional layers have
3x3x3 kernel then ReLU, that produce a decreasing number of feature
maps from $192$ to $12$.

Multiclass classification (6 tissues and background), is obtained with
a last layer of convolution with 7 1x1x1 filters, followed by a softmax 
activation.

\section{Experiments}
\label{sec:experiments}

We qualitatively compared the segmentation pipelines of FSL,
FreeSurfer, SPM, ANTs and the 3D U-Net on T1-weighted images from
patients with agenesis of corpus callosum (ACC) and with complex
distortions of brain anatomy, as described in Section
\ref{sec:distorted}. Different pipelines produced different sets of
segmented tissues. To harmonize the results, we considered only 3 main
tissues: gray matter (GM), white matter (WM) and the cerebrospinal
fluid (CSF). These tissues are the ones of greater interest in most of
the applications of brain tissue segmentation. Moreover, for the
pipelines segmenting the deep gray matter (DGM), we labeled DGM as
GM. The technical details of each pipeline are the following:
\begin{itemize}
\item \textbf{FSL v6} \cite{jenkinson_fsl_2012}: we used
  \texttt{fsl\_anat}\footnote{\url{https://fsl.fmrib.ox.ac.uk/fsl/fslwiki/fsl_anat}}
  with default parameter values.
\item \textbf{FreeSurfer v6} \cite{dale_cortical_1999}: we used
  \texttt{recon-all}\footnote{\url{https://surfer.nmr.mgh.harvard.edu/fswiki/recon-all}}
  with default parameter values.
\item \textbf{SPM12} \cite{ashburner_voxel-based_2000}: we used default
parameter values and default prior.
\item \textbf{ANTs v2.2.0} \cite{tustison_large-scale_2014}: we used
  \texttt{AntsCorticalThickness.sh} with default parameter vales and
  the PTBP (pediatric) prior \cite{avants_pediatric_2015}. We
  considered only the steps till tissue segmentation.
\item \textbf{3D U-Net.} The model was implemented with TensorFlow
  1.8.0 \cite{martin_abadi_tensorflow_2015} trained on T1-weighted
  images from the Normal Brains Datasets of $\approx 900$ images
  described in Section \ref{sec:materials}. We kept apart 90 randomly
  selected images for a sanity check, i.e., to assess that the trained
  model could reach state-of-the-art quality of segmentation on
  healthy subjects. The training (loss: cross-entropy, Adam
  optimization, learning rate: $10^{-4}$) was performed iteratively,
  one image at a time taken at random from the training set, looping
  over the whole set for 60 epochs.
\end{itemize}
All computations were performed on a dedicated workstation: 6 cores
processor Intel(R) Xeon(R) CPU E5-1650 v4 3.60GHz, 128Gb RAM, GPU:
NVIDIA GeForce GTX 1080TI 11Gb RAM.

\subsection{Results}
\label{sec:results}
Except for FreeSurfer (\texttt{recon-all}), all pipelines carried out
all the required segmentations. FreeSurfer failed in all cases of
severe anatomical distortion. Specifically, \texttt{recon-all} did not
converge during either Talairach registration or skull stripping.
SPM completed the segmentation task in all subjects. However, its
confidence in the prior for the segmentation initialization and
optimization leads to many major macroscopic errors. Given the strong
limits on the length of this article, we do not illustrate and discuss
the uninteresting results of FreeSurfer and SPM.

\begin{figure}
  \centering
  \includegraphics[width=13cm]{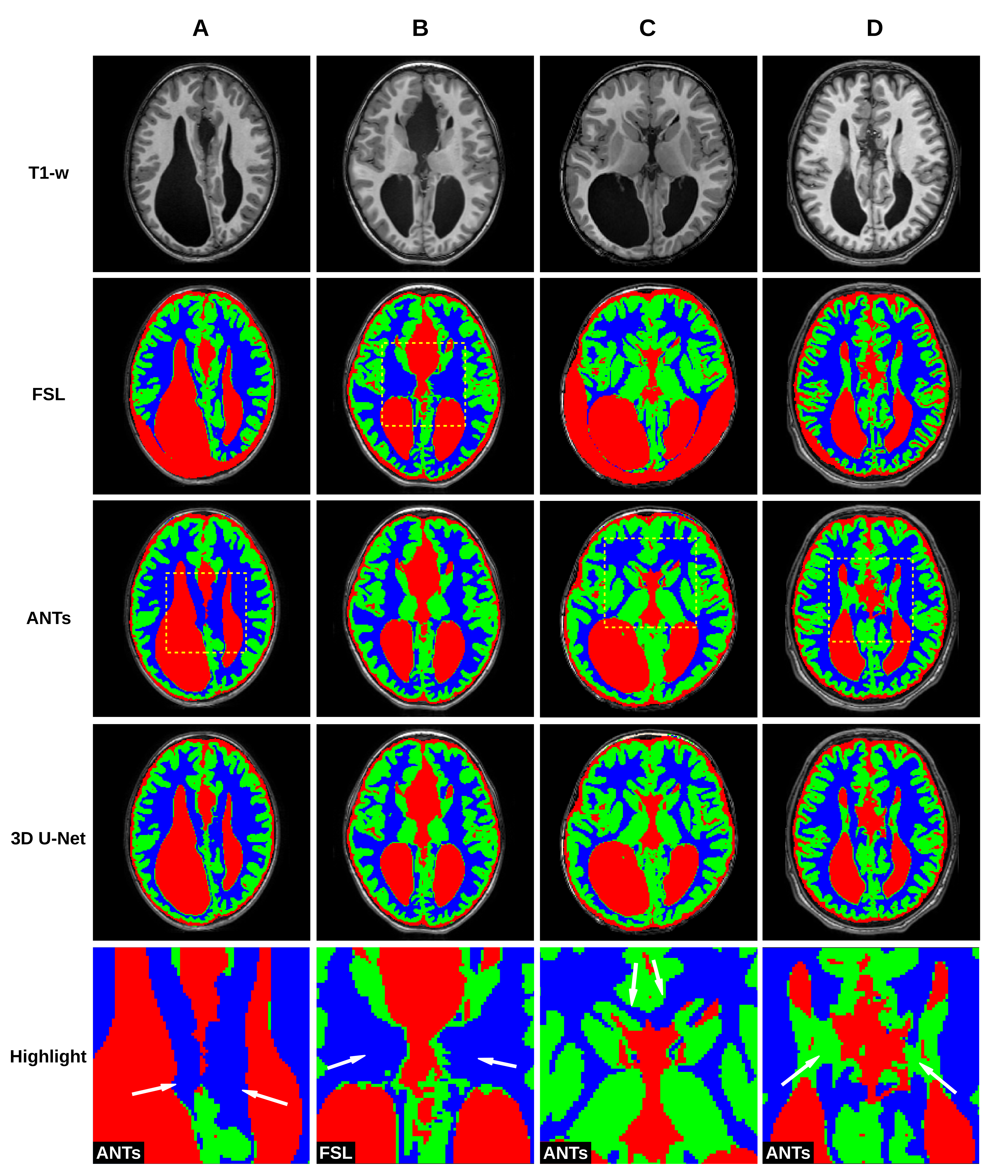}
  \caption{First row: T1-weighted MR images of 4 subjects (A, B, C and
    D) with \emph{agenesis of corpus callosum}. Below, the related
    tissue segmentations (GM in green, WM in blue and CSF in red) of
    the following pipelines: FSL (2nd row), ANTs (3rd row) and 3D
    U-Net (4th row). In the 5th row, for each subject, we show the
    enlarged view of one of the segmentations, indicated above with a
    dashed yellow square. White arrows point to the highlights
    discussed in Section \ref{sec:discussion}.}
  \label{fig:acc}
\end{figure}

\begin{figure}
  \centering
  \includegraphics[width=13cm]{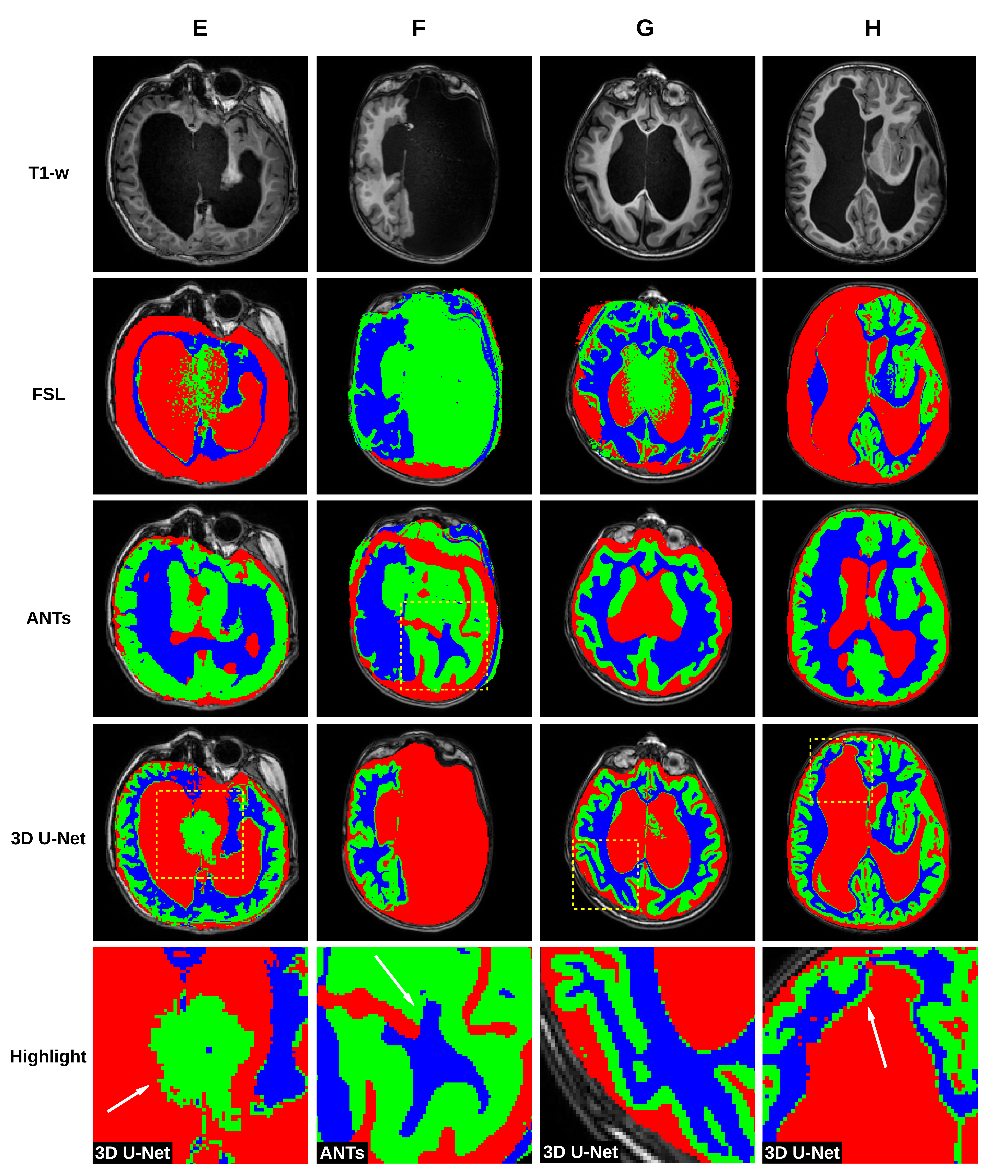}
  \caption{First row: T1-weighted MR images of 4 subjects (E, F, G and
    H) with \emph{complex cerebral distortions}. Below, the related
    tissue segmentations (GM in green, WM in blue and CSF in red) of
    the following pipelines: FSL (2nd row), ANTs (3rd row) and 3D
    U-Net (4th row). In the 5th row, for each subject, we show the
    enlarged view of one of the segmentations, indicated above with a
    dashed yellow square. White arrows point to the highlights
    discussed in Section \ref{sec:discussion}.}
  \label{fig:complex}
\end{figure}

In Figure \ref{fig:acc}, we show a paradigmatic set of axial slices
segmented by FSL, ANTs and the 3D U-Net, from patients with ACC,
i.e. from 4 of the 12 subjects described in Section
\ref{sec:distorted}. Similarly, in Figure \ref{fig:complex}, we report
paradigmatic axial slices segmented by those methods, from 4 of the
subjects with complex cerebral distortions. For the segmentations of
all methods, we re-labeled as CSF all the voxels inside the brain mask
of each patient that were incorrectly segmented as background. In both
figures in the last row, for each subject, we highlight a detail of
the slice for one of the segmentation methods (indicated with a dashed
square, above in the same column). Such details are discussed in
Section \ref{sec:discussion}.

Finally, in Table \ref{tab:results_healthy}, we report the results of
the sanity check, i.e. that the training process of the 3D U-Net on
normal brains was successful. The numbers represent the average
quality of segmentation (DSC, Dice similarity coefficient) obtained by
the 3D U-Net for the reference segmentation of the 6 tissues of normal
brains described in Section \ref{sec:preprocessing}. These results are
comparable to those in the state-of-the-art
\cite{spalletta_convolutional_2018,yogananda_deep_2019}.

\begin{table}
  \centering
  \begin{tabular}{p{1.0cm} | p{1.4cm} | p{1.43cm} | p{1.4cm} | p{1.4cm} | p{1.4cm} | p{1.4cm} | p{2cm}}
    Metric              &    CSF &     GM &     WM &    DGM &  Trunk
    & Cereb. & \textbf{Grand Avg.} \\
    \hline
    DSC	 & 0.87$\pm$0.06 & 0.95$\pm$0.04 & 0.95$\pm$0.03 & 0.93$\pm$0.02 & 0.93$\pm$0.03 & 0.96$\pm$0.02 & 0.93$\pm$0.03  \\
    \hline
  \end{tabular}
  \caption{3D U-Net: average Dice similarity coefficient (DSC) for
    segmentations of 6 tissues, plus grand average, on 90 normal
    brains, after 60 epochs of training.}
  \label{tab:results_healthy}
\end{table}

\section{Discussion}
\label{sec:discussion}

Figure \ref{fig:acc} shows that FSL fails to segment GM and WM in
cases of moderate and severe ventricular dilatation with thinning of
the WM (case A, C and less evident in D). In one case (B) FSL also
misses identifying the thalami as (deep) GM. ANTs performs better in
identifying (deep) GM and the cortex at the convexity. However, case A
and C show that it may fail in differentiate between GM and
subcortical WM on the mesial surface of the hemispheres. This may be
related to the prior used by that pipeline, which is based on the
anatomy of normal subjects and not designed to recognize spatial
reorganization of the cortex, especially in the midline, like in ACC
cases. A similar error is in case C where a cortical component close
to the head of the caudate is misclassified as WM. Finally in D,
Probst bundles, which are abnormal WM tracts running parallel to the
medial ventricular wall, are labelled as GM. In contrast, the 3D U-Net
performs well in segmenting ACC. The most relevant error in these
cases is at the interface between ventricles and WM: the 3D U-Net
wrongly identifies a very thin layer of GM along the inner ventricular
surface. This is probably related to partial volume effects.

Figure \ref{fig:complex} shows that, in case of complex malformations
and severe parenchymal distortion, FSL and ANTs are unreliable and
incur in major macroscopic errors, as opposed to 3D U-Net which
performs vastly better. In cases of severe ventricular dilatation and
distortion (E and H), FSL fails to segment the cortex, which is
wrongly labelled as CSF. In case F, the CSF collection that replaces
the left hemisphere is misclassified as the cortex. In G, FSL fails to
properly segment the pachygyric (i.e. with a reduced number of gyri)
cortex. Finally, some intensity inhomogeneities in the deep
ventricular CSF are misclassified as GM (E and G). With ANTs, which is
based on priors, the pipeline is forced to segment WM, GM and CSF in
the missing hemisphere (F) or when the anatomy is highly irregular. In
all of these cases, the pipeline misplaces structures (E, G, H) or
segment structures that are actually missing (F). The 3D U-Net
outperforms FSL and ANTs in all cases (e.g. G), with few mistakes. The
main issues are: i) the mislabeling of signal inhomogeneities in the
deep CSF (E, same as FSL), and the segmentation of a subtle layer of
GM at the border between lateral ventricles and WM (H, as done in ACC
cases). Care must be used in evaluating CSF/WM interface in cases of
brain malformations because, at this level, heterotopic GM nodule may
occur.

\subsection{Conclusions}
\label{sec:conclusions}
In this work, we observe a much higher accuracy of the 3D U-Net when
segmenting brains with different degrees of anatomical distortion,
compared to well-established pipelines. This is surprising given that
such cases were not used in the training phase. At the same time, the
3D U-Net can reproduce the high quality of segmentation of ANTs on
normal brains. Clearly, the results on distorted brains are not
perfect but still a valuable starting point for manual refinement by
experts/neuroradiologists. In future work we plan to manually segment
T1-w images of the distorted brains, to create a gold standard and to
be able to quantify the quality of segmentation of the 3D U-Net and to
use some of them during the training process of the network.

\section{Acknowledgment}
\label{sec:acknowledgment}

Data used in the preparation of this article were obtained from the
C-MIND Data Repository created by the C-MIND study of Normal Brain
Development. This is a multisite, longitudinal study of typically
developing children from ages newborn through young adulthood
conducted by Cincinnati Children's Hospital Medical Center and UCLA
and supported by the National Institute of Child Health and Human
Development (Contract \#s HHSN275200900018C). A listing of the
participating sites and a complete listing of the study investigators
can be found at \url{https://research.cchmc.org/c-mind}. This
manuscript reflects the views of the authors and may not reflect the
opinions or views of the NIH.

\clearpage

\bibliographystyle{splncs03}

\bibliography{nilab_malformation}

\section*{Supplementary Materials}

\begin{center}
  \includegraphics[width=12.4cm]{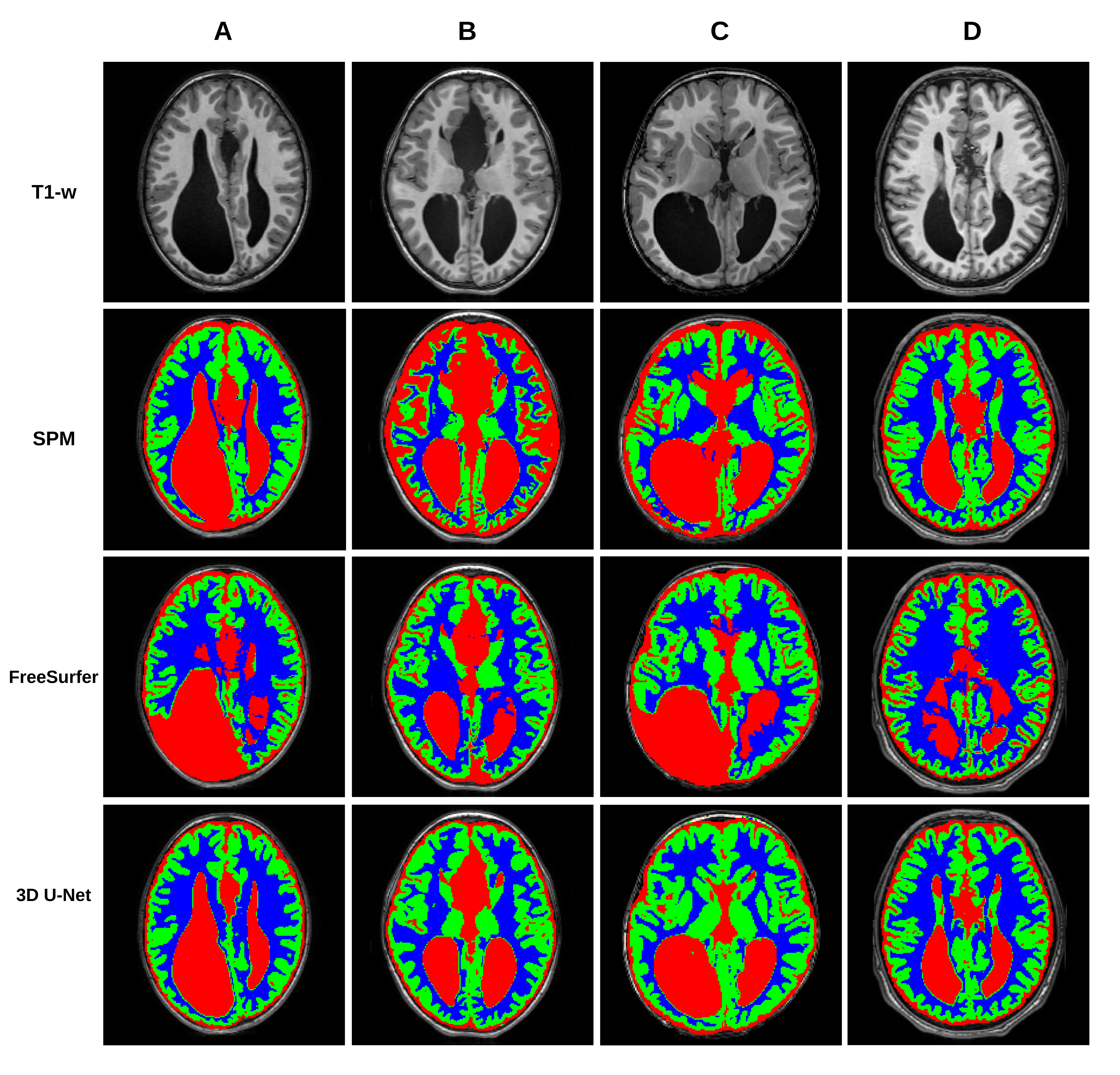}
\end{center}
This is an extension of Figure 1, i.e. paradigmatic segmentations of
patients with Agenesis of Corpus Callosum, where we show SPM and
FreeSurfer that we excluded in the manuscript, together with the 3D
U-Net for comparison. The segmentations of SPM and FreeSurfer have
major errors in the ventricles and in the area of the thalami and
caudate. SPM probability maps were thresholded at $p=0.4$ which was
the optimal choice to reduce overlap between tissues and to reduce
false positives.

\begin{center}
  \includegraphics[width=12.4cm]{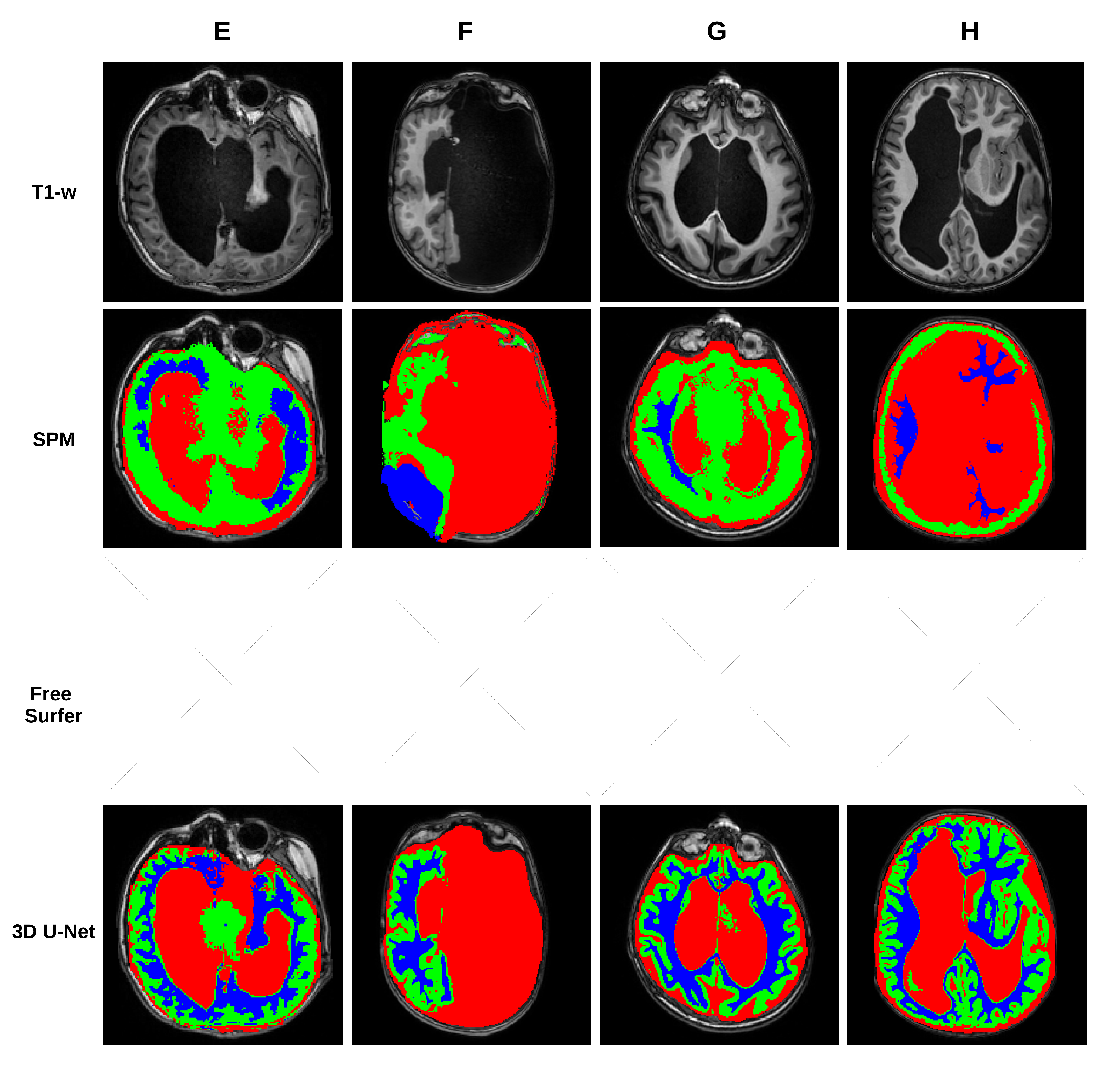}
\end{center}
This is an extension of Figure 2, i.e. paradigmatic segmentations of
patients with complex cerebral distortions, where we show SPM and
FreeSurfer that we excluded in the manuscript, together with the 3D
U-Net for comparison. SPM has major macroscopic errors almost
everywhere. FreeSurfer failed to converge on the images of these
subjects so the related entries are empty, as reported in Section
4. SPM probability maps were thresholded at $p=0.4$ which was the
optimal choice to reduce overlap between tissues and to reduce false
positives.

\end{document}